\documentclass[aps,prb,superscriptaddress,a4paper,twocolumn]{revtex4-1}

\usepackage{amsmath}
\usepackage{color}
\usepackage{amssymb}
\usepackage{graphicx}

\begin{document}
\title{On the martensitic transformation in Fe$_{x}$Mn$_{80-x}$Co$_{10}$Cr$_{10}$ high-entropy alloy}

\author{P. Singh}\email{psingh84@ameslab.gov}
\affiliation{Ames Laboratory, U.S. Department of Energy, Iowa State University, Ames, Iowa 50011 USA}
\author{S. Picak}.
\affiliation{Dept. of Mechanical Engineering, Texas A\&M University, College Station, TX 77843, USA}
\affiliation{Dept. of Materials Science \& Engineering, Texas A\&M University, College Station, TX 77843, USA}
\author{A. Sharma}
\affiliation{Sandvik Coromant R\&D, Stockholm, 12679 Sweden}
\affiliation{Ames Laboratory, U.S. Department of Energy, Iowa State University, Ames, Iowa 50011 USA}
\author{Y.I. Chumlyakov}
\affiliation{Tomsk State University, Siberian Physical Technical Institute, Novosobornay Square 1, 634050 Tomsk, Russia}
\author{R. Arroyave}
\affiliation{Dept. of Mechanical Engineering, Texas A\&M University, College Station, TX 77843, USA}
\affiliation{Dept. of Materials Science \& Engineering, Texas A\&M University, College Station, TX 77843, USA}
\author{I. Karaman}
\affiliation{Dept. of Mechanical Engineering, Texas A\&M University, College Station, TX 77843, USA}
\affiliation{Dept. of Materials Science \& Engineering, Texas A\&M University, College Station, TX 77843, USA}
\author{Duane D. Johnson}\email{ddj@iastate.edu, ddj@ameslab.gov}
\affiliation{Ames Laboratory, U.S. Department of Energy, Iowa State University, Ames, Iowa 50011 USA}
\affiliation{Dept. of Materials Science \& Engineering, Iowa State University, Ames, Iowa 50011 USA}

\begin{abstract} 
High-entropy alloys (HEAs),  and even medium-entropy alloys (MEAs), are an intriguing class of materials in that structure and property relations can be controlled via alloying and chemical disorder over wide ranges in the composition space. Employing density-functional theory combined with the coherent-potential approximation to  average over all chemical configurations, we tune free energies between face-centered-cubic (fcc) and hexagonal-close-packed (hcp) phases in Fe$_{x}$Mn$_{80-x}$Co$_{10}$Cr$_{10}$ systems.~Within Fe-Mn-based alloys, we show that the martensitic transformation and chemical short-range order directly correlate with the fcc-hcp energy difference and stacking-fault energies, which are in quantitative agreement with recent experiments on a $x$=40~at.\% polycrystalline HEA/MEA.  Our predictions are further confirmed by single-crystal measurements on a$x$=40at.\% using transmission-electron microscopy, selective-area diffraction, and electron-backscattered-diffraction mapping. The results herein offer an understanding of transformation-induced/twinning-induced plasticity (TRIP/TWIP) in this class of HEAs and a design guide for controlling the physics behind the TRIP effect at the electronic level.
\end{abstract}

\maketitle

{\par }High-entropy alloys (HEAs) \cite{1,MIRACLE2017448,George2019} and medium-entropy alloys (MEAs) are new exciting class of materials  with their vast design space and emerging unique properties \cite{2,3,4,IKEDA2019464,57}. Originally, single-phase solid-solution formation in HEAs was proposed to originate through entropy maximization \cite{1}, but recent evidence \cite{5,6,7,8} suggest that entropic description is oversimplified, as HEAs/MEAs may display significant local chemical short-range order \cite{9,10,PhysRevMaterials.4.113802,Zhang2020}. Outside the HEA space, high-Mn (15-30 at.\%) austenitic ($\gamma$-fcc phase) ferrous alloys are a material class that receive special attention due to their low/medium stacking fault energies (SFEs) \cite{15,PhysRevMaterials.3.113603}, high ultimate tensile strength ($>$ 1000 MPa) with total elongation over 60\% at room temperature (RT) \cite{16}, and uses in automotive industry \cite{17}. A number of studies were performed to tune SFEs in the austenitic alloys to achieve better control over governing deformation mechanisms, e.g., dislocation slip ($\ge$40 mJ/m$^{2}$), mechanical twinning (20-40 mJ/m$^{2}$), twinning-induced plasticity, and/or martensitic transformation ($\le$20 mJ/m$^{2}$), including those in HEAs \cite{19,20,21,66}. As expected, the deformation at the low-SFE regime in austenitic ($\gamma$-fcc) alloys is mainly realized through a martensitic ($\epsilon$-hcp) transformation \cite{22}. The fcc-based, single-phase multi-principal-element solid-solutions have gained attention due to their outstanding ductility, however, lower strength limits their use for engineering applications \cite{HE20149,SCHUH2015258}. Recent work has shown that a martensitic transformation \cite{3, Lu2018, PhysRevLett.122.075502} or precipitation strengthening \cite{ZHAO201772, Yang933} could provide an effective way to address the strength-ductility trade-off in this important materials class.

{\par}Here we employed density-functional theory (DFT) methods to understand how to tune some of these key properties via alloying and disorder, in particular, formation-energy (E$_{form}$), stacking-fault energy (SFE), and short-range order (SRO) of fcc-based solid-solutions, see supplement for details on methods \cite{27,28,29,30,31,32,33,34,35,36,37,38}. We show that  chemistry profoundly alters E$_{form}$ and SFE of Fe$_{x}$Mn$_{80-x}$Co$_{10}$Cr$_{10}$  and can suppress SRO and, hence, long-range order (LRO). The DFT calculated E$_{form}$ and SFE, together with experimental observations reveal details of the strain-driven martensitic (fcc$\rightarrow$hcp) transformation at  $x$=40at.\%Fe. Our DFT-predicted SRO \cite{9} on the $x$=40~at.\%Fe system  indicate very weak chemical SRO and, hence, very low-temperature ordering behavior (below 50 K). The predicted low-temperature ordering suggests the preference for forming martensite rather than long-range order. Molecular-dynamics (MD) simulations \cite{PLIMPTON19951} on this system at RT also show a strain-driven martensitic transformation at 40~at.\% Fe. The results reveal key underpinning of physical principles behind formation of martensite, and an opportunity for more intelligent design of high-performance HEAs \cite{26} -- for a more directed exploration of higher-dimensional composition space \cite{4}.

\begin{figure}[t]
\includegraphics[scale=0.23]{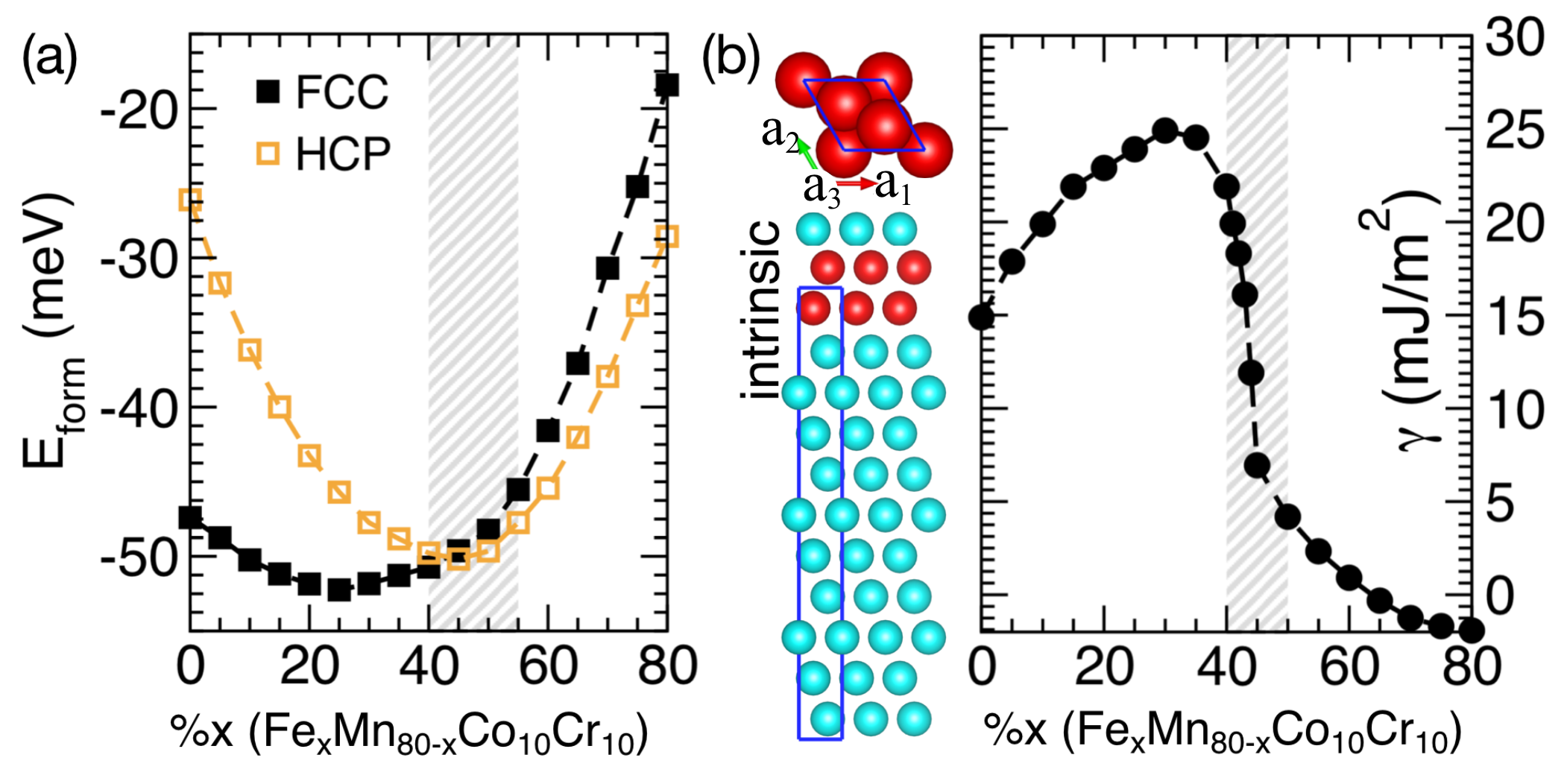}
\caption{(Color online) For Fe$_{x}$Mn$_{80-x}$Co$_{10}$Cr$_{10}$ ($x$=0-80\%), (a) formation energy (E$_{form}$ in meV), see Fig.~S1 for volume vs. $x$, and  (b)  intrinsic SFE (mJ/m$^2$), plus a schematic of stacking sequence and outlined unit cell (11-sites). In (a), the dual-phase (fcc+hcp) region is shaded, where SFE (b) drops rapidly with increasing \%Fe.}
\label{fig1}
\end{figure}

{\par}Following Hume-Rothery, phases stability of HEA/MEA systems in different lattice structures can be estimated empirically using valence-electron count (VEC), e.g., bcc (e.g., A2 or Laves phase) for VEC$<$7; coexistence of bcc/fcc at 7$<$VEC$<$8; and fcc for VEC$>$8 \cite{4}. The solute and host with similar VEC show large solubility, i.e., a metal dissolve one of higher valency to a greater extent or lower valency to a lesser extent. These critical values can be directly and more reliably evaluated using DFT \cite{39}. Notably, Mn-based fcc alloys are known for lower VEC than empirically defined solid-solution phase limit, the exceptions are already noted in \cite{39,41}. As such, DFT calculations were performed on Fe$_{x}$Mn$_{80-x}$Co$_{10}$Cr$_{10}$ to avoid  limitations of empirical rules and detail the thermodynamic stability and planar faults versus  $x$ (Fig.~\ref{fig1}).

{\par} Phase stability (Fig.~\ref{fig1}a) shows a critical range of 40-55 at.\%Fe with a possible two-phase region at the crossover in the stability of the fcc and hcp phases, with fcc energetically favorable at low \%Fe. The dual-phase alloy in the Fe-rich region can benefit from solid-solution strengthening, owing to the decreased SFE \cite{3,42}. The SFE for Fe$_{x}$Mn$_{80-x}$Co$_{10}$Cr$_{10}$ (Fig.~\ref{fig1}b)  first increases with \%Fe additions and then shows a  precipitous drop within the dual-phase region from 30--50 at.\%Fe. The low SFE indicates an fcc lattice for this system becomes unstable with respect to the formation of  intrinsic stacking faults. These results correlate well with Fe-Mn phase diagram that shows $\epsilon$-martensite is not formed above 50 at.\% Fe, and higher Mn concentration drives  austenite-only structure \cite{43}. In a pioneering work, Kelly investigated Fe-Ni and Fe-Ni-C alloys with relatively high SFE \cite{44} and showed that alloys with appreciable Cr or Mn have low SFE and form martensites associated with planar stacking faults or the formation of `hcp' martensite \cite{45,46}. 

{\par}Typically, medium or negative SFE has been regarded as the crucial indicator of TRIP phenomena \cite{49,50,51,Li2017,Huang2018,CHEN2020138661,LI2017323}, suggesting a low-energy barrier for  fcc-to-hcp transformation \cite{52}. Conventional steels are known for martensitic transitions at medium SFEs, e.g., Fe-Mn-based alloys with SFEs below 11-19 mJ/m$^{2}$, and Co-Ni-Cr-Mo alloys below 9-15 mJ/m$^{2}$ \cite{53}. This has also been exemplified for Cu-Al \cite{54}, Ni-Cu \cite{55}, Ni-Fe \cite{55}, Ni-Co \cite{56}, Co-Ni-Cr-Mo \cite{56}, and Fe-Mn based (TWIP/TRIP steel) alloys \cite{16, 42}. However, the difficulty in measuring SFE \cite{15,49,50,51} makes comparison with theory harder. In Table~\ref{tab1}, we tabulated calculated SFEs for Fe$_{x}$Mn$_{80-x}$Co$_{10}$Cr$_{10}$ with a comparison to experiments \cite{57} and other Fe-Mn-based alloys \cite{42,59,60}. Our calculations indicate medium-to-low intrinsic SFE for Fe$_{x}$Mn$_{80-x}$Co$_{10}$Cr$_{10}$ with increasing \%Fe. Extrinsic (ESF) and twinning (TFE) fault energies (Fig.~S2) suggests intrinsic faults are energetically more favorable, i.e., ISFE $<$ ESFE $<$ TFE. In Fig.~\ref{fig1}b, the ISFE is non-monotonic versus $x$, where the energy needed to alter the fcc stacking sequence \cite{61} is varying dramatically and goes negative with at.\%Fe with composition. {The dramatic change in ISFE can be attributed to the relatively large increase in hcp volume compared to fcc (see red zone in Fe-rich region in Fig.~S1). The negative ISFEs in fcc configurations suggest that the hcp stacking would be preferred energetically.}  ESFEs (Fig.~S2d) follow a similar trend as ISFE, but unlike ISFE it remains positive in the Fe-rich region. No such composition dependence versus \%Fe was observed in TFE (Fig.~S2d).

\begin{table}[t]
\centering
\begin{tabular}{cc||ccc}\hline
                        {\bf  Fe-Mn-based}         &                            &                  &{\bf Fe$_{x}$Mn$_{80-x}$Co$_{10}$Cr$_{10}$} \\ \hline
                 {\bf Systems}                        &  {\bf SFE}             & {\bf \%x}   & {\bf SFE}       &\\ \hline
                                                               &                             &                  & Theory & Expt   \\ \hline
Fe$_{66}$Mn$_{28}$Al$_{3}$Si$_{3}$ &   38.8$\pm$5        & 0              &  14.7   & --\\
Fe$_{69}$Mn$_{25}$Al$_{3}$Si$_{3}$ &   21.0$\pm$3        & 10            &  19.8   & --\\
Fe$_{72}$Mn$_{22}$Al$_{3}$Si$_{3}$ &  15.0$\pm$ 3        & 20            &  21.9   & --\\
Fe$_{75}$Mn$_{25}$                            &   27.5$\pm$ 3.3    & 40            & 22.3    & 17$\pm$4\\
Fe$_{78}$Mn$_{22}$                            &  15.0$\pm$ 1.8     & 45            &   7.3    & --\\
Fe$_{80}$Mn$_{20}$                            &  18.0$\pm$ 2.2     & 60            &   1.1    & --\\
Fe$_{82}$Mn$_{18}$                            &  22.0$\pm$ 2.6     & 70            & -1.3     & --\\
Fe$_{84}$Mn$_{16}$                            &  26.0$\pm$ 3.1     & 80            & -1.9     & --\\
\hline 
\hline
\end{tabular}
\caption{For Fe$_{x}$Mn$_{80-x}$Co$_{10}$Cr$_{10}$, DFT-calculated SFE at 0 K with comparison to experiments \cite{57} at 300 K and Fe-Mn-based alloys \cite{15,42,59,60}.}
\label{tab1}
\end{table}

\begin{figure}[t]
\includegraphics[scale=0.4]{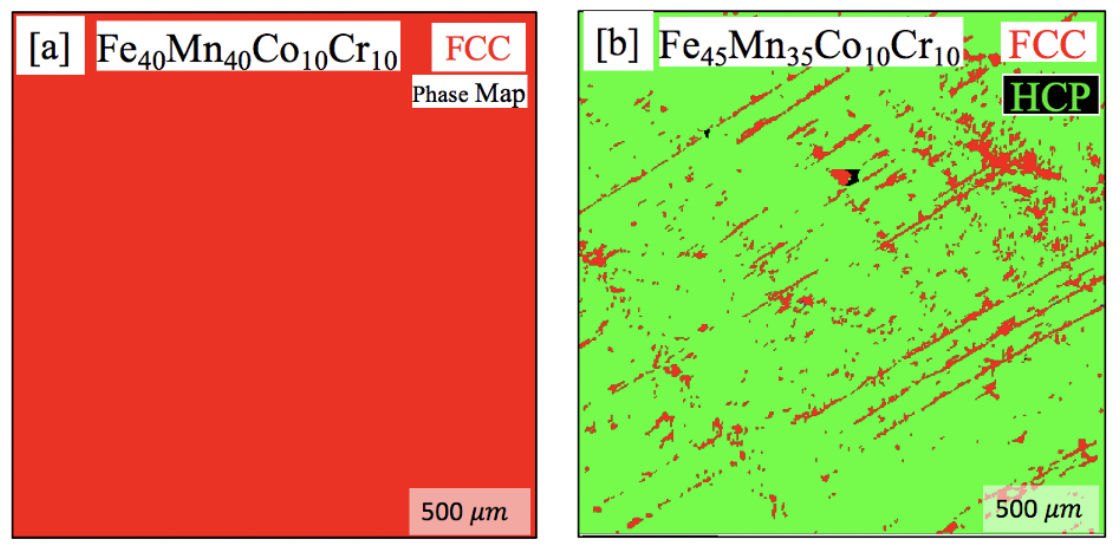}
\caption{(Color online). EBSD phase maps of (a) 40~at.\%Fe, and (b) 45~at.\%Fe alloys show single-phase (fcc) and dual-phase (fcc+hcp) microstructure, respectively.}
\label{fig2}
\end{figure}

{\par}Phase stability analysis of Fe$_{x}$Mn$_{80-x}$Co$_{10}$Cr$_{10}$ in Fig.~\ref{fig1}a shows dual-phase region with onset at 40 at.\%Fe. Recently, the $x$=50~at.\%Fe alloy has been reported as a two-phase at RT \cite{3}, whereas the $x$=40~at.\%Fe alloy is single-phase fcc at RT \cite{57}. DFT results (Fig.~\ref{fig1}) show small $\Delta{E}_{form}^{fcc-hcp}$ for 40 and 45~at.\%Fe, with a higher SFE for 40~at.\%Fe (22.2 mJ/m$^{2}$) compared to  45~at.\%Fe (7.3 mJ/m$^{2}$). Thus, the higher SFE of fcc 40~at.\%Fe alloy plays a key role in stabilizing the single-phase fcc; that is, RT cannot provide enough thermal energy to drive the martensitic  transformation in contrast to 45~at.\%Fe. To prove our claim, we grew a 40 and 45~at.\%Fe single-crystal HEAs (see experimental methods in supplemental information). Electron backscatter diffraction (EBSD) micrographs (Fig.~\ref{fig2}a,b) show single-phase (fcc) and dual-phase (fcc+hcp) microstructure, respectively, for 40 and 45~at.\%Fe.

{\par} With proper configurational averaging for general HEAs (using the coherent-potential approximation, not just one representative configuration), theory provides a reliable and quantitative prediction of $\Delta{E}_{form}^{fcc-hcp}(x)$ and ${\gamma}_{SFE}(x)$ and its dramatic composition dependence, here for Fe$_{x}$Mn$_{80-x}$Co$_{10}$Cr$_{10}$. The small $\Delta{E}_{form}^{fcc-hcp}$ and medium SFE at $40$~at.\%Fe provide crucial theory guidance  for the microstructural evolution in HEA steels. 

\begin{figure}[t]
\includegraphics[scale=0.35]{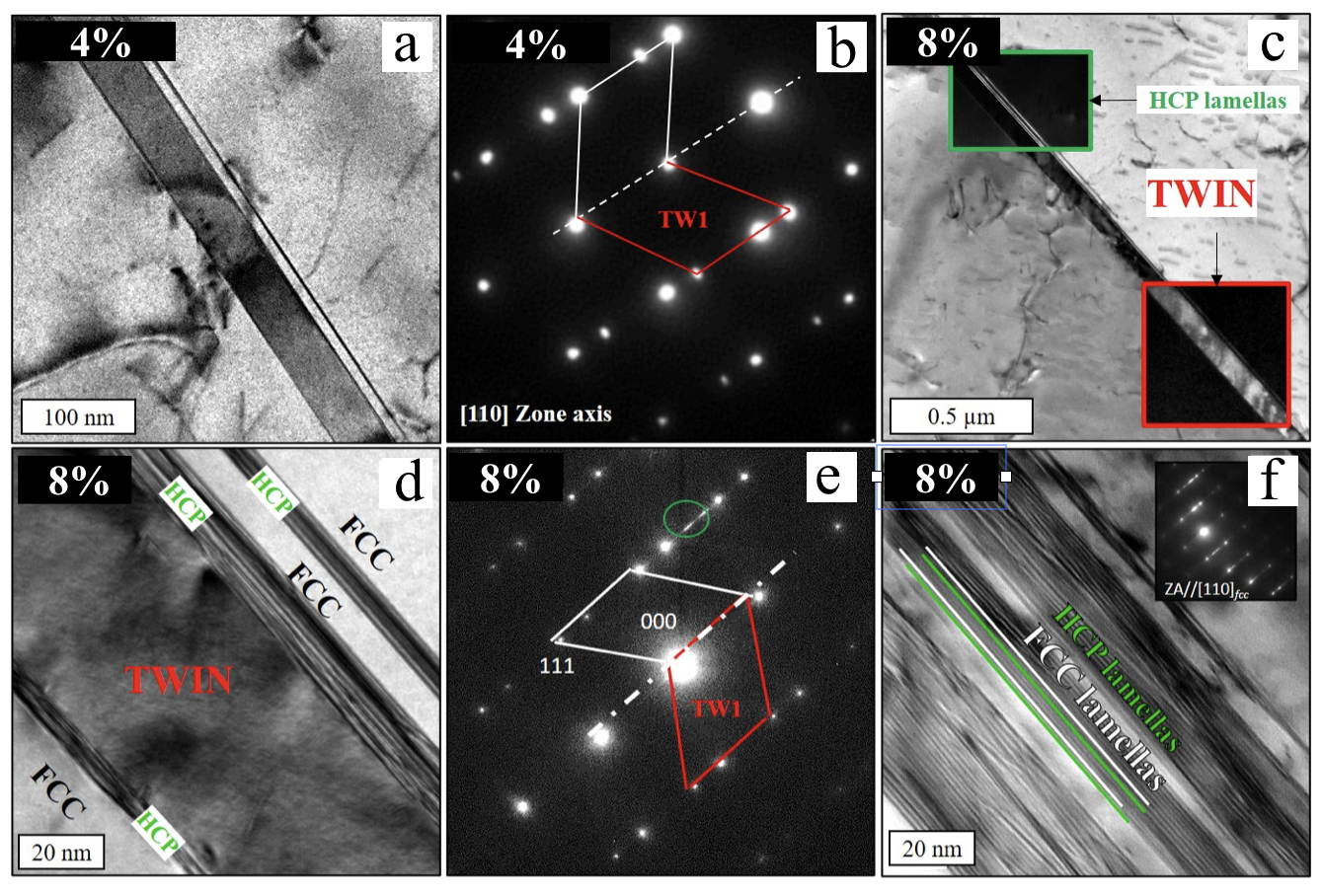}
\caption{{(Color online). Bright-field and dark-field TEM micrographs and SADP of [111]-oriented single-crystal with $x$=$40$~at.\% exhibiting $\epsilon$-martensitic transformation and twin nucleation. (a) Nano twins at 4\% strain, and (b) corresponding SADP. (c) Nucleation of $\epsilon$-martensite at the twin boundary at 8\% strain (inset - dark-field images confirms this), (d) higher magnification of (c), and (e) corresponding SADP. (f) Nano $\epsilon$-martensite/fcc bundles.}}
\label{fig3}
\end{figure}

{\par} To confirm, we examined 40~at.\%Fe single crystals under RT uniaxial tensile loading. The bright-field TEM micrographs and selective-area diffraction patterns (SADP) are shown in Fig.~\ref{fig3} (also see Fig. S5). {At 4\% strain, nano-twin formation was observed in Fig.~\ref{fig3}a at the beginning of deformation, which is confirmed by SADP in Fig.~\ref{fig3}b. With further increase in strain, $\epsilon-$martensite was activated at strains as low as 8\% at twin boundaries in Fig.~\ref{fig3}c. The strain-induced martensitic transformation in Fig.~\ref{fig3}d and corresponding SADP at Fig.~\ref{fig3}e at higher magnification further confirms the role of competing fcc/hcp stability with medium SFE (Fig.~\ref{fig1}a). X-ray measurements in Fig.~S6  further confirm the martensite (hcp phase) is not an artifact of TEM thin-foil effect. Nano-sized hcp and fcc lamellas in Fig.~\ref{fig3}f reveal a composite microstructure acting as a barrier for the dislocation motion, which improved the strain hardening behavior (see Fig.~S7). Recent reports also confirm that simultaneous activation of TWIP/TRIP effect provides better strength and ductility combination \cite{PICAK2021113995}.}

{\par}The connection between SRO and low-temperature ordering behavior is very important for alloy design \cite{9,23}. Upon cooling, the high-temperature disordered phase gives rise to SRO and ultimately at low-temperature to ordering. And, SRO in the disordered phase is often a `precursor' to the long-range order at low temperatures (or competition between local ordering and clustering). The Warren-Cowley SRO pair-correlations $\alpha_{\mu\nu}^{ss'}({\bf k};T)$ were calculated directly using thermodynamic linear-response theory -- for more details see \cite{9,23,24,37,38,39,40}. Here, $s,s'$  indices denote sublattices in a crystal structure [1 (2)  for fcc (hcp)] and $\mu,\nu$ denote elements [here 1--4]. For   N-component solid-solutions, all $\frac{1}{2} N(N-1)$ SRO pair correlations (arising from fluctuations in site-occupation probabilities) are calculated simultaneously \cite{9}, similar to that done to get the vibrational stability matrix (i.e., phonon modes and ``force-constants''). SRO  is dictated by pair-interchange energies (chemical stability matrix), i.e., S$_{\mu\nu}^{ss'}({\bf k};T)$ \cite{9,23}; the thermodynamically-averaged second-variation of the free energy with respect to compositional fluctuations \cite{9}. As such, the most unstable SRO mode with wavevector {\bf k}$_o$ will have the largest peak in $\alpha_{\mu\nu}^{ss'}({\bf k}_o;T>T_{sp})$ for a specific $\mu$-$\nu$ pair in the solid-solution. An absolute instability to ${\bf k}_o$ mode \cite{9} occurs below the spinodal temperature $T_{sp}$, where $[\alpha_{\mu\nu}^{ss'}]^{-1}({\bf k}_o;T_{sp})=0$. If $k_{o}$=(000), the alloy is unstable to segregation. Both ordering and clustering peaks may compete.  Importantly, S$_{\mu\nu}^{ss'}({\bf k}_o;T)$ dictates the origin for the SRO, which may be a different pair that peaks in the observable $\alpha_{\mu\nu}^{ss'}({\bf k}_o;T)$, as they are related exactly by an inverse in linear response, see \cite{4,9,24,38}.

\begin{figure}[t]
\includegraphics[scale=0.25]{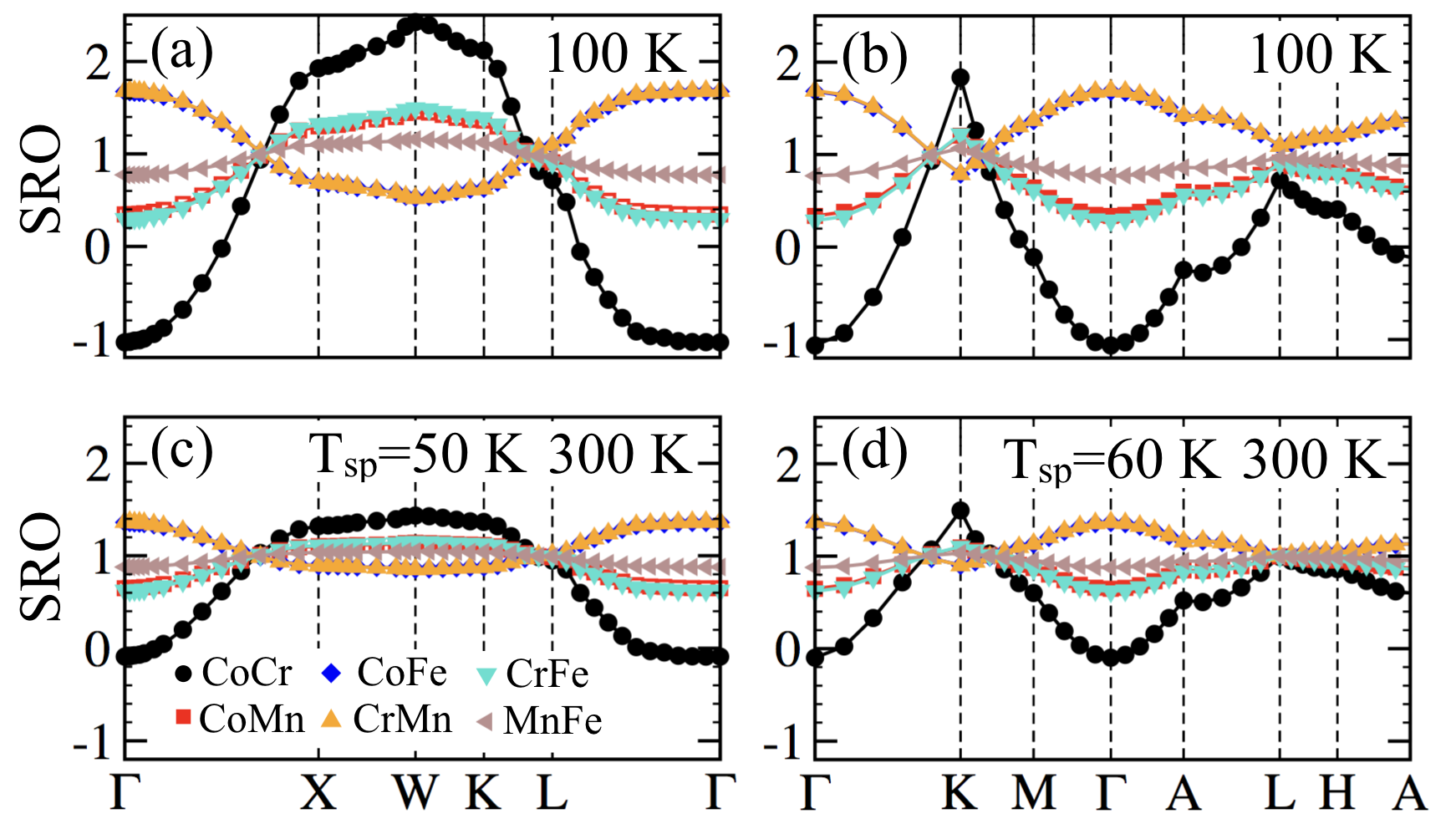}
\caption{(Color online). Warren-Cowley SRO  $\alpha_{\mu\nu}^{ss'}({\bf k};T)$ parameters in Laue units for 40~at.\%Fe at  (a,b) 100 K, and (c,d) 300 K plotted along fcc (a,c) and hcp (b,d) high-symmetry  Brillouin zones directions, respectively.}
\label{fig4}
\end{figure}

{\par}For the 40~at.\%Fe alloy, the calculated SRO for fcc and hcp phases are shown in Fig.~\ref{fig4} at 100 K and 300 K (RT). Although SRO at RT is not strong relative to 100 K, it  persists over a range of temperature, which may impact dislocation glide, as found in fcc solid-solution alloys~\cite{Zhang2020}. The spinodal decomposition in solid-solutions occurs during, e.g., order-disorder transformation during cooling \cite{63}, in which the spinodal temperature indicates the absolute instability to the ${\b k}_{o}$ mode in SRO \cite{9}. A high mixing entropy keeps solid-solution phases stable at higher temperatures, becoming metastable at low temperatures. {To estimate temperature changes on relative stability, we approximate free-energy $\Delta{F}$[fcc-hcp] by including SRO and electronic entropy as $\Delta{F} = \Delta{E}_{form} - T(\Delta{S}_{SRO} + \Delta{S}_{elec})$ in fcc and hcp phases with respect to high-T disorder fcc phase (1500 K having no SRO) with $\Delta{F}$=$-1.25$ meV/atom; with lowering of temperature,  $\Delta{F}$[SRO] = $-2.0$ meV/atom at 1000 K and  $-8.4$ meV/atom at 300 K. Although  the energy of fcc lowers relative to  hcp, the change is weak. Therefore, no major impact is expected for transition temperature.  
At low temperatures (with SRO included), the increased stability of fcc over hcp further conforms with the experimentally observed single-phase fcc at 300 K, see Fig.~\ref{fig2}a.} The calculated T$_{sp}$ for the fcc and hcp $40$~at.\%Fe alloy is 50 and 60 K, respectively. Such low phase-decomposition temperatures indicate that it cannot be retained at RT, supporting a martensitic transformation, as predicted in Fig.~\ref{fig1} and observed in Fig.~\ref{fig3}.

{\par} Importantly, the state of local chemical SRO is often a precursor to low-temperature order through cooling in most alloys. Notably, the microstructure and local order both can be controlled by composition and/or heat treatment at higher temperatures. To reveal the local chemical order at the onset of dual-phase in Fig.~\ref{fig1}a), we analyze the Warren-Cowley SRO parameters $\alpha_{\mu\nu}^{ss'}({\bf k};T)$ that manifest the observable diffuse intensities at 100 K and 300 K for $40$~at.\%Fe. The diffuse intensities in Fig.~\ref{fig4}a-d  have maximal SRO at W=$(1\frac{1}{2}0)$ in fcc phase (indicating D0$_{22}$-type ordering) and at K=$(\frac{2}{3}\frac{2}{3}0)$ for hcp phase (indicating D0$_{19}$-type order), which have possible origin in peaks in S$_{\mu\nu}^{ss'}$  \cite{9} (see Fig.~S8 that shows weak temperature dependence). The Co-Cr pair is the most dominant mode that become unstable at T$_{sp}$ of 50 K in fcc and of 60 K hcp phase. The Cr-Mn and Co-Fe pairs contribute with the second most dominant modes with peaks at $\Gamma=(000)$ both in fcc and hcp phases at RT. The presence of SRO at RT, however weak, can impact dislocation glide \cite{10}. Our bright-field TEM image (Fig.~S3) shows dislocations pile-up at the onset of plastic deformation, indicating strongly localized dislocation structures along a specific (111) planes in fcc alloy. {According to Cohen and Fine \cite{Cohen1962}, the first dislocation in the pile-up is exposed to higher resistance against slip due to interaction with the favorable (stable) SRO environment, which leads to localized deformation and pile-up in Fe$_{40}$Mn$_{40}$Co$_{10}$Cr$_{10}$. All successive dislocations, produced by the activated dislocation source and moving along the regions with SRO that was locally destroyed (due to rearrangement of solute) help to overcome the higher resistance, which subsequently helps to nucleate the martensitic phase during deformation. The small $\Delta{E}_{form}^{fcc-hcp}$ poses only a small athermal transformation energy barrier between fcc and hcp phases that further assists the strain-induced martensitic transformation in Fig.~\ref{fig3}.} Thus, the dislocation behavior observed (Fig.~S3) and martensitic transformation shown in Fig.~\ref{fig3} can be associated to the weak SRO, similar to binaries \cite{47,51}, as SFE and high yield strength are already known to have a minor effect on the dislocation pileup \cite{Zhang2020}.

\begin{figure}[t]
\includegraphics[scale=0.45]{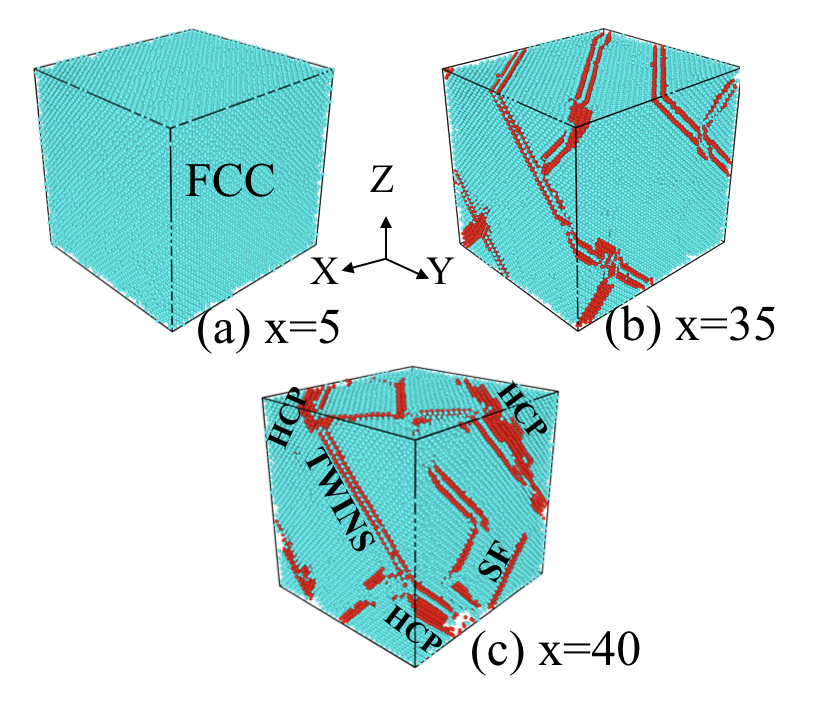}
\caption{(Color online). (a-c) Stress-induced martensitic formation from MD  under uniaxial (111) loading at 8\% strain in Fe$_{x}$Mn$_{80-x}$Co$_{10}$Cr$_{10}$. Twinnability increases with \%Fe, with martensite at x=40 at.\%. Mirrored single-layer (red atoms) indicate twinboundaries and two hcp layers within the fcc structure (cyan atoms) signify intrinsic stacking-fault (SF).}
\label{fig5}
\end{figure}

{Finally, MD simulations were performed (Fig.~\ref{fig5}) to understand the deformation mechanism in 40~at.\%Fe under uniaxial stress with increasing strain at RT (method and details in supplement). The microstructure of uniaxially-deformed Fe$_{x}$Mn$_{80-x}$Co$_{10}$Cr$_{10}$ at 8\% strain (matching experimental condition) enhanced the ability to form SFs and twins (TWs) with increase in at.\%Fe, as shown in Fig.~\ref{fig5}. Our deformation analysis suggests that (see movie in supplement Video~S1) intrinsic fault planes act as source for twin nucleation. Smallman {\it et al.}~\cite{SMALLMAN1964145} also discussed that lower SFE is preferable for twins as it helps to accommodate large strain, see \cite{TADMOR20042507,Kibey2007}. This mechanism becomes important as, unlike high SFE materials, low SFE alloys cannot develop cross slips that helps to absorb large stress. Once the deformation twins are formed further increase in strain can either increase TW density or existing twins act as nucleation sites for the hcp, especially at intersection of SFs and TWs, see Fig.~\ref{fig3}c-d.   Shockley partials were found as the primary dislocations during early loading stage. While Hirth dislocations and stair-rods, identified at the later stages, can be responsible for enhancement in strength and ductility due to the formation of Lomer-Cottrell lock (see movie in supplement Video~S1), which agrees with an extra-stage strain-hardening observed in stress-strain curve for 40~at.\%Fe (see Fig.~S7).}

In conclusion, using DFT-based Green's function methods in combination with proper configuration-averaging using the coherent-potential approximation, we predicted the controlling physics behind the martensitic transformation in a complex Fe$_{x}$Mn$_{8-x}$Co$_{10}$Cr$_{10}$ solid-solution alloy system to occur at the specific composition of $x$=$40$~at.\%Fe. We confirmed the theoretical predictions using precision experiments on single-crystal samples. Molecular dynamics simulations supports both DFT prediction and our experimental observation of a martensitic transformation. The tunability of phase energy and stacking-fault energy in HEAs/MEAs using purely chemistry and disorder shows the relevance of theory-guided design for the next-generation alloys with superior structure-property correlation, as well as the unique insights for controlling phase transformation in technologically relevant alloys.

\section*{Acknowledgements}
PS and SP contributed equally to this work. Research at Ames Laboratory was supported by the U.S. Department of Energy (DOE), Office of Science, Basic Energy Sciences, Materials Science \& Engineering Division. Ames Laboratory is operated by Iowa State University for the U.S. DOE under contract DE-AC02-07CH11358. RA acknowledges the support of QNRF under Project No. NPRP11S-1203-170056.

\bibliography{Manuscript_Final}
\bibliographystyle{rsc}

\end{document}